# CFD SIMULATIONS FOR HYDRAULIC SHORT-CIRCUIT FEASIBILITY ANALYSIS

A. Morabito, C. Wu, S. Sigali, E. Vagnoni

**Abstract:** Hydraulic Short Circuit (HSC) application allows the simultaneous pumping and generating operations on different units of the same pumped hydro energy storage (PHES) plants for the extension of the power consumption range. This article proposes a methodology based on time-history statistics and CFD numerical simulations to investigate the feasibility in integrating such a kind of operation in pre-existing hydropower plants. The evaluation of the potential power output and pressure drop analysis in the new HSC operation are presented. The penstock trifurcation of a selected PHES plant is also analysed by CFD simulations, by evaluating the loss-head coefficients and comparing the flow conditions in different PHES operations.

## 1 Introduction

Pumped hydro energy storage (PHES) plants have been adopted in the last decades to mitigate the installation boost of intermittent energy sources, such as wind and photovoltaic power systems [1]. At the same time, recent economics and policy trends pinpoint the challenge of high flexibility in both power generation and storage modes. In PHES, such operating control, which could be awarded by ancillary service retributions depending on the local energy market regulations, are partially achieved in generation mode with the power adjustment provided by the Francis turbine guide vanes or by the Pelton turbine spear valve and deflectors. However, in pumping mode there is no possibility to manage the power consumption because the guide vanes in the pump-turbine usually stay fully open [2, 3].
Exceptionally, variable speed and/or seldom geometry regulations allow PHES plants to extend their operating flexibility. Nevertheless, for already existing PHES, the economic justifications of adding variable speed capacity might be somewhat poorer especially to small-hydro projects [4].
Hydraulic short circuit (HSC) configuration, corresponding to the simultaneous operation of the pumps and turbines, enhances the power consumption flexibility of PHES. The main advantage of this practice is regulating the net absorbed energy by the PHES plant with same power regulation range equal to that of the turbine operation [2]. If the hydropower system is equipped with ternary configuration (or with multiple reversible units), the HSC can be obtained and supply primary and secondary frequency regulation services to the transmission system operators (TSOs) within a larger marketable capacity. The power regulation in HSC provided by reversable units are inherently lower than in ternary solutions. The hydraulic pump-turbine cannot operate simultaneously in both modes, and it has a smaller range of operability than turbines, especially for Pelton turbines.

Today, the HSC principle counts a very few active applications in the world, but academic and industrial communities have increasingly devoted fervent interest in the profitability of this emerging operation with optimal market participation and reserve scheduling [5, 6, 7, 8].

Upgrading the existing hydropower plant infrastructures for HSC is not an easy task and comprehensive feasibility study is required. About the resiliency and reliability of HSC applications, Nicolet [9] analyzed the contribution of a PHES to compensating the consumer load fluctuations and wind power output variation, together with a thermal power plant, in a mixed islanded power network. Also, Perez-Diaz et al. [10] confirmed the adding value of HSC-PHES to the load-frequency regulation of an isolated system from the perspective of both the TSO and power plant owner.

In PHES operating in HSC modes, the flow derivation and power losses in the penstock connected between upper reservoir and more than the units is complex, and the accurate evaluation of the power compensation is crucial in assisting the power plant operation. Kong [11] proposed a method to calculate the head loss in a shared tunnel for a PSHP with variable speed pumps but it doubly overestimates the loss, whereas HSC scheme, in fact, reduces the power loss in the shared tunnel. Skjelbred [12] proposed a mathematical formulation to calculate the actual loss in the shared tunnel when a PSHP operates in HSC mode and the presented method is effective to offset the overrated loss and obtain the correct result. Computational fluid dynamics (CFD) simulations are carried out to obtain more details of the water flow structures during HSC within the hydraulic circuit. Huber et al. [13] conducted CFD simulations to investigate the flow characteristics and head losses of a T-junction planned in the Kopswerk II under turbine, pumping and three different HSC operating conditions. An alternative T- junction was simulated to prove a better computed design in flow properties and lower head-losses. Decaix et al. [14] focus on one of the junctions of the Grand-Maison power plant that is recently upgraded to run in HSC mode: CFD studies have been carried out for various ratio of the flow discharge between the pumps and the turbines to determine the pressure losses in the HSC modes.

This paper contributes to the definition of the methodology in the HSC feasibility analysis and relating the effect of such operations to a PHES trifurcation. First, the potential of HSC integration to a selected PHES is discussed assessing the limits of the power regulation (Section 3). In Section 4, the CFD analysis approach is presented and applied to two different scenarios of HSC operations involving two turbomachine units. Section 5 provides the numerical results head loss coefficients defining the pressure drop relative to the branches towards the turbine and the upper reservoir. Moreover, this paper leaps from other works in the literature by presenting the effect of the inlet velocity profile to the trifurcation pressure drop. Finally, the conclusions are given in Section 6.

## 2   Research objectives

This research article aims to investigate the feasibility of HSC integration by CFD analysis of the hydropower plant trifurcation. HSC operation was not foreseen when the selected PHES was designed. The pressure losses through the trifurcation must be estimated and considered for defining the power-net output. Flow patterns and pressure distributions in the penstock determine the actual head losses. Fluid-dynamic behaviour analysis is needed to diagnose the health state and behaviour of all the essential components of the hydro-mechanical system.

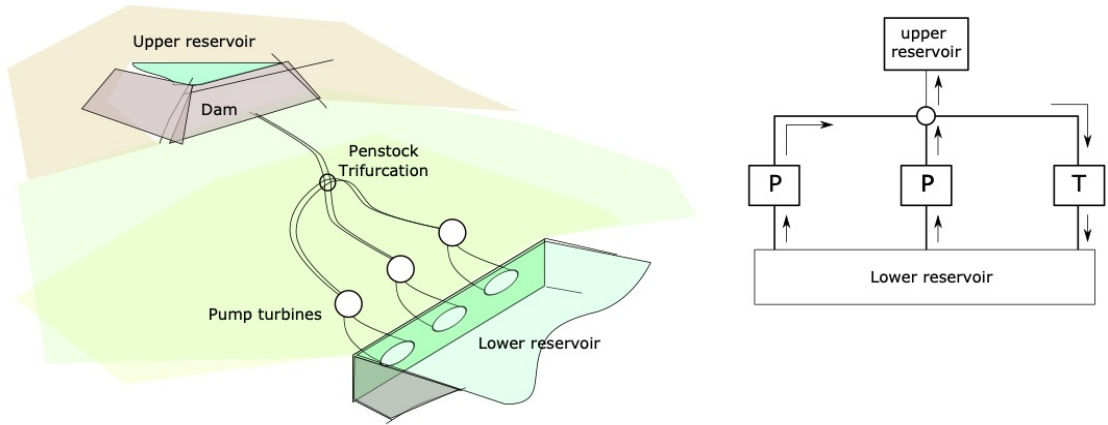

Fig. 1: A schematic view of the investigated PHES plant and position of the three units. On the right, two pumps (P) and a single turbine (T) are operating in HSC configuration.

Penstock elements include transitions, bends, tees, elbows, and reducers. They are especially susceptible to excessive vibrations, aging, and lining loss [3]. Trifurcations can be found in the waterways between the headwater or tailwater stretches and the hydropower unit. Their task is dividing or unifying water flow with minimum losses. In HSC operations, variable discharge per branch has a significant impact in the head losses in such kind of connections. Based on their design, location and punctual operating condition, complex flow patterns may occur.

To learn as much as possible about these circumstances and HSC feasibility in existing PHES plants, this work analyses the hydraulic performance and the stability of the penstock trifurcations of the selected case study for a safe operation under such unusual conditions. To support the HSC feasibility assessment, depicting limits and operational configurations, CFD numerical simulations are performed to study the flow behaviour of the PHES waterways in the critical sections of the energy storage plant.

## 3 HSC potential in the selected case study

This research enhances the quantification of HSC potential for an existing PHES plant commissioned in the early '70s, featuring three alike Francis pump-turbines (Fig 1). Concerning power management and dispatching strategy, three operations are finally available in a modern PHES power plant: generating mode, pumping mode and HSC. Figure 2 illustrates the different operating cases with their relative flexibility for the selected PHES system under head (H) variation. In generating mode, the turbines can operate under variable discharge (Q) coming from the upper reservoir to the lower reservoir (Q>0) and inject power to the grid (P>0). Turbines benefit from the variable guide vanes openings to regulate inlet flow to the runner and, thus, the angular momentum at the shaft. When the unit operates in pumping mode, water is delivered in the upper reservoir through the fix-speed pump (Q<0). Pumps generally work at a single operating condition defined by a fixed Q-H characteristic curve under constant rotational speed, because inlet guide vanes are not common in pumps. Lastly, HSC operation region can quickly be obtained on the same reference chart of Fig. 2 by translating the turbine characteristic origin at the pump duty point [Hp, Pp]. The turbine gross head is provided from the pump units and a fraction, or the totality of the pump discharge is delivered to the turbines while the rest is pumped to the headwater. The demarcation of this portion is set to the targeting PHES power consumption serving the power consumption flexibility. Provisioning water both from the pumping units and the upper reservoir is conceivable when necessary.

Preliminary study on the time-history statistics of the studied PHES shows that the power consumption range in HSC mode is extended in the bounds of 5-70% of the nominal pumping consumption when a single unit is running as pump and one as turbine. In particular, HSC tackles the extension of the operating range targeting an additional continuous power consumption from 92% to 173% of one pump nominal power, by engaging two machines in pumping mode and one in generating mode (HSC with three units running).

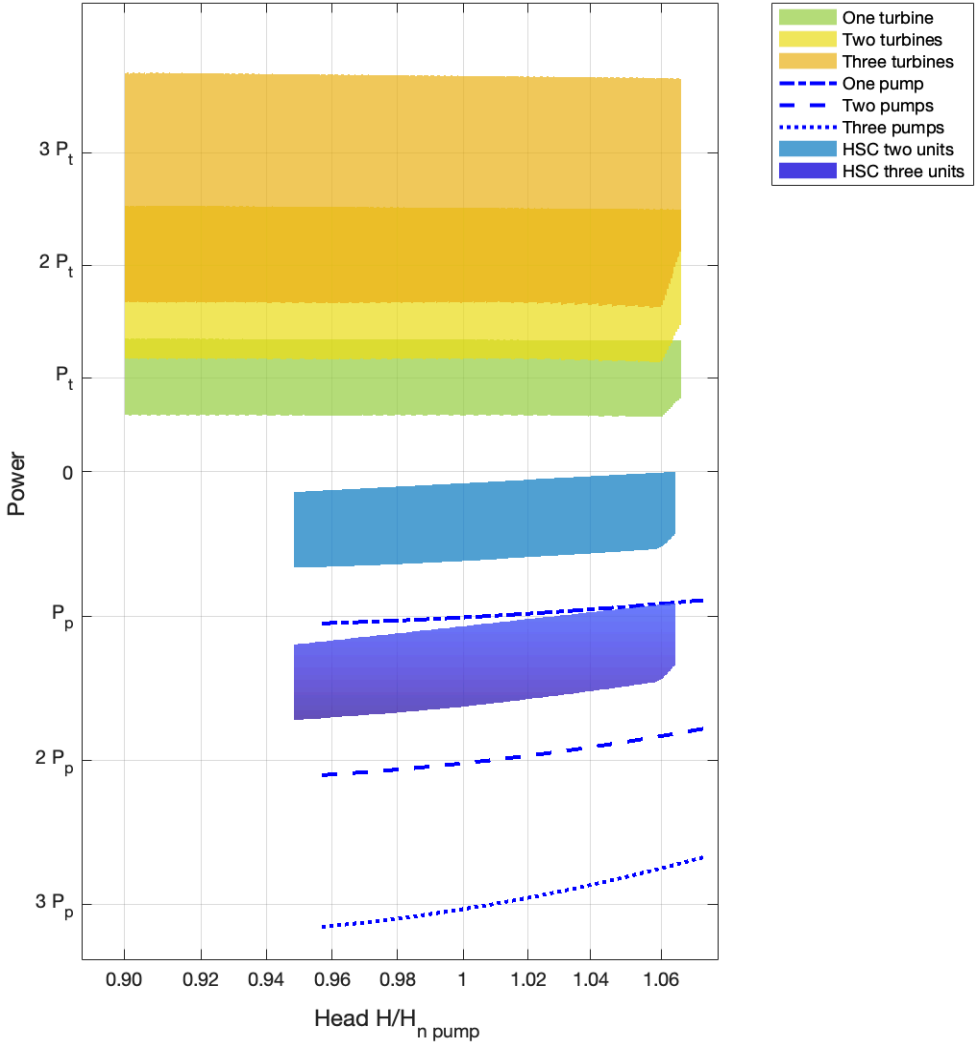

Fig. 2: Power regulation chart of the PHES for generating, pumping and HSC modes.

# 4 Numerical simulation strategy

## 4.1 Overview

In the following sections the flow simulations setup is presented. The scope of numerical simulation analysis is to compare the steady state regimes of a single unit in generating mode with HSC operation with one unit in pumping and another in generating mode.

| Case   | # unit 1      | # unit 2      | # unit 3        | Discharge        | Power         |
|--------|---------------|---------------|-----------------|------------------|---------------|
| Case-1 | Pumping mode  | Off           | Generating mode | $(0.5 - 1.1)\,Q_p$ | $-P_p + P_t$ |
| Case-2 | Off           | Pumping mode  | Generating mode | $(0.5 - 1.1)\,Q_p$ | $-P_p + P_t$ |

Table 1: Simulated operation of the trifurcation hydropower plant

## 4.2 Geometric model description

The trifurcation geometry has been obtained by a 3D scan on-site and reproduced typically in CATIA V5. The three branches toward the hydraulic units have a final diameter of 2 m and the penstock's diameter is 5 m large. The two external branches bend of 50° angle to meet at the trifurcation with 8 m inner radius curvature. At the crossing sections, two stiffeners up to 0,6 m tall surround the connection of the central branch (Fig. 3).

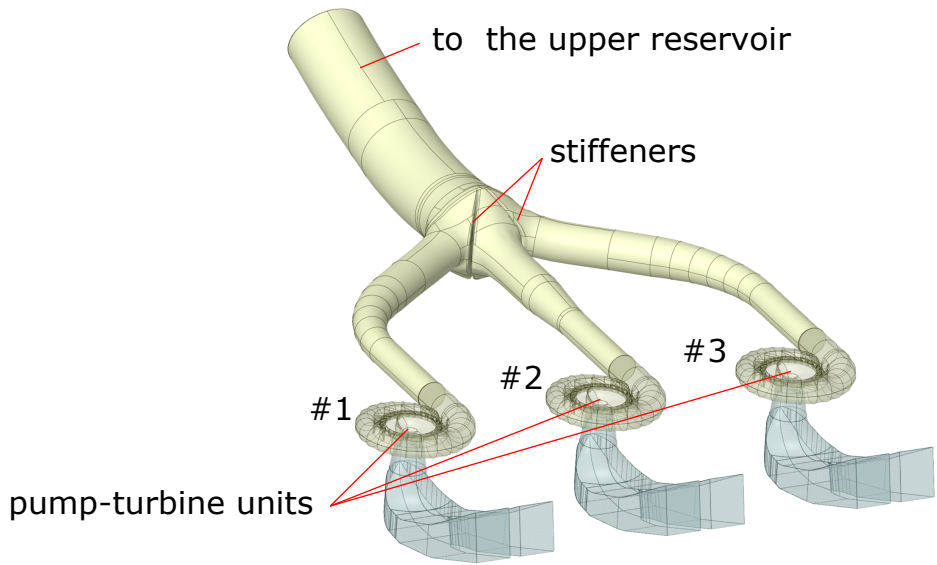

Fig. 3: View of the wet surfaces of the trifurcation at the PHES plant.

## 4.3 Grid generation methodology

The generated geometry is then imported to Ansys Fluent Mesh 2021R2 for meshing and polyhedral meshes are generated for discretizing computational fluid domain. Grid independency tests are performed to exclude effects of the mesh size on the numerical solutions. Finer local sizing is implemented in the proximity of critical areas as in nearabout of the stiffeners. In addition, an inflation mesh technique is used for the fluid volume near the pipe wall, which generates multiple thin layers of elements within the boundary layer region to better capture the near wall flow performance. The size of the first cell at the wall is always checked to agree with the adopted turbulence model. A grid independency study is carried out monitoring the variation of the averaged pressure in control sections for converged solutions by the increasing number of grid nodes. To ease the convergence of the numerical calculations, the outlets are extended by a length of ten times the diameter, avoiding possible recirculation and hard convergence.

### 4.4 Numerical methods

All simulations are performed by the CFD software Ansys Fluent 2021R2 with the *realizable k – ε* turbulence model. The y+ values are consistent with the model adopted, laying in the range of 30-300 for all the simulated discharge ranges in the involved domain regions.
In enhanced wall treatments, each wall-adjacent cell's centroid should be located within the log-law layer, 30< y+ < 300 requires a greater first cell height, thus, a reduced total number of nodes than a standard model (y+ = 1) [4]. The well-known SIMPLEC scheme is chosen to deal with the pressure-velocity coupling and second order upwind spatial discretization scheme is employed for momentum, turbulent kinetic energy and turbulent dissipation rate. The inlet boundary is set as velocity inlet while the outlet boundaries to the upper reservoir and turbine are set as mass flow outlet and pressure outlet respectively. No-slip boundary condition is adopted for all the physical walls. Cases under various HSC operation conditions can be obtained by changing the discharge of outlet connected with upper reservoir. In those cases where the solution appears to fluctuate, under-relaxation is activated for the pressure and turbulence factors. The convergence criterion is not defined unequivocally by examining the residuals, but also from the imbalance of mass and momentum and stable outputs [4].

## 5 Results and discussions

### 5.1 Pressure losses analysis

The pressure losses in the trifurcation are calculated over fixed sections of the penstock to monitor the progressive total pressure drop. The loss coefficient (or the resistant coefficient) is defined as follow [5]:

$$K_L = h_L / (V^2 / 2g)$$

where $h_L$ is the pressure drop in meters and V is the averaged flow speed at the rated section as in V [m·s$^{-1}$] = Q [m$^3$·s$^{-1}$] / A [m$^2$]. Fig. 4 illustrates the head loss coefficients in the trifurcation of the flow delivery in the upper reservoir $\Delta P_u$ and the turbine branch $\Delta P_t$ during the HSC operations. The results are showed over the discharge ratio in the two branches: Qt / Qp = 0 the whole pump discharge is delivered to the upper reservoir; Qt / Qp = 1 the whole pump discharge feeds the turbine. Fig. 4-a represents the case-1, namely the HSC configuration involving the units 1, 3 running. The loss coefficient Kt presents downward trend when Qt/Qp range from 0.1 to 0.3 and show upward trend when Qt/Qp is greater than 0.3. Total flow losses in penstock mainly accounts for viscous dissipation of internal secondary flow and wall friction losses. Under lower discharge rate (Qt/Qp = [0.1 - 0.3]), the flow velocity in the turbine branch is low in respect to the nominal condition. The increase of the discharge leads to the decrease of secondary flow losses which causes a reduction of the loss coefficient. By increasing the discharge ratio, wall friction losses increase and cause the sharp increase of losses in the turbine branch. On the upper reservoir side, the wall friction factor does not significantly affect the totality of the losses because the flow velocity remains low. The penstock is designed for accepting the discharge from all the units simultaneously without overburdening the PHES hydraulic losses.

Fig. 4-b illustrates the pressure drop of the trifurcation for the case-2 where unit 2 is running as pump and unit 3 as turbine. The pressure losses in the turbine branch show upward trend with the increase of the discharge ratio because of the increase of wall friction losses which is caused by the increasing flow velocity. For the upper reservoir branch, losses caused by viscous dissipation of secondary flow accounts for most of the flow losses by considering the lower flow velocity in this branch which explains why the value of Ku shows upward trend with the turbine discharge.

Compared with Kt, Ku is lower which results from lower flow losses caused by secondary flow and wall friction due to the lower flow velocity and smaller flow deflection in the upper reservoir branch. As shown in Fig.4, Kt and Ku of case-1 are higher than in the case-2, due to additional flow losses related to the different geometry of the branch connected to the pump.

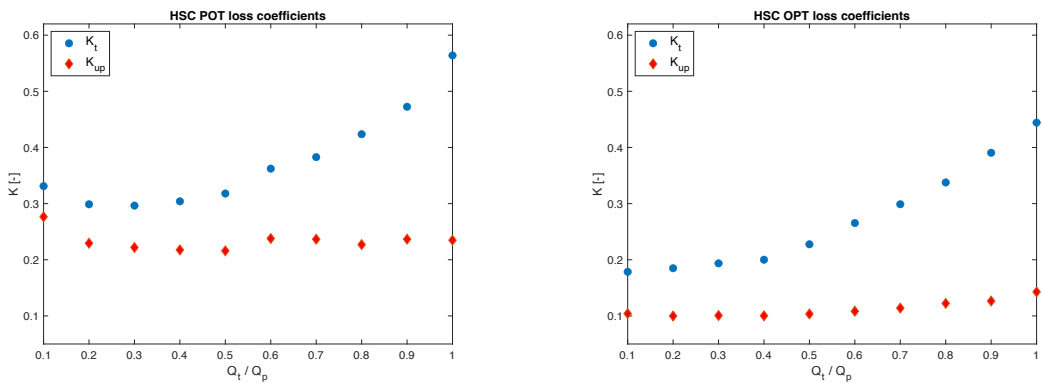

Fig. 4: (a), head loss coefficient for case-1    (b), head loss coefficient for case-2

Interestingly, the resulting pressure and velocity fields of the trifurcation are affected in the HSC operations, in which the inlet boundary conditions are not imposed as uniform velocity profile but extracted from the 3D numerical simulations of the pump-turbine unit running in pumping mode. The three velocity components at the outlet of the spiral case are imported in the solver for the trifurcation analysis as new inlet boundary condition. Fig. 5 illustrates a detail of the computed pump geometry and the velocity profile contours of the pump outlet. The CFD simulations of the pump-turbine is not here discussed as out of scope of this article.

To define the effect of the computed velocity profile at the inlet of the domain, the evolution of the total pressure is traced by several sections and grouped in three portions of the trifurcation: the pumping branch $\Delta P1$ (from section S6-S4), the intersection $\Delta P2$ (between sections S4-S3) and, the branch to the turbine $\Delta P3$ (S3-S0).

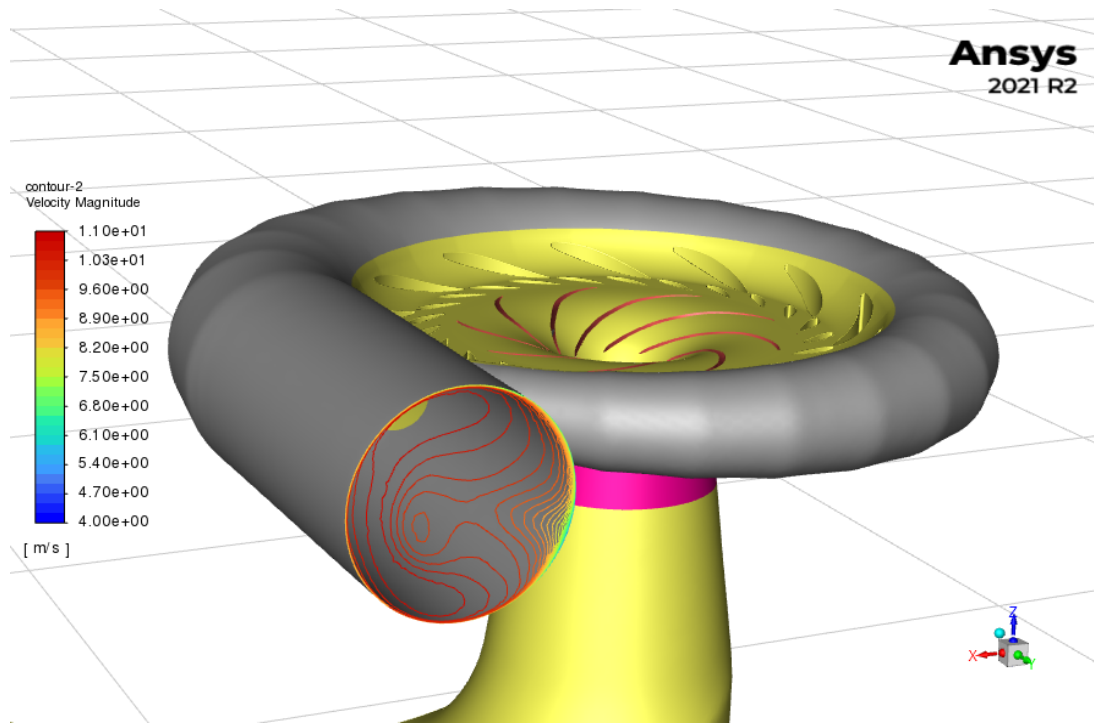

Fig. 5: Illustration of the pump-turbine and the CFD velocity contours at the outlet of the pump spiral casing.

The head loss portion and totality, ΔPtot = ΔP1+ΔP2+ΔP3, are normalized to the maximum recorded value of each case. The analysed operating conditions space from the minimum discharge required by the turbine $Q_{t,\,min}$ up to the maximum, $Q_{t,\,max}$. The latest operating point prompts additional discharge from the upper reservoir to fulfil with the required turbine discharge.

### 5.1.1 HSC case 1

The inlet velocity profile change does not considerably impact the pressure drop in the ΔP1 and ΔP3 portions, pump and turbine branches respectively (Fig. 6) and both reach a similar maximum value. The discharge is invariant in the pump branch while the turbine discharge spaces from its minimum and maximum, by generating a constant relative pressure drop ΔP1 of 20% over the maximum value recorded for this operating condition. On the other side, ΔP3 raises with the discharge that engages the turbine branch to finally reach a value similar to ΔP1. The highest local pressure drop is located at the crossing point of the trifurcation for the section expansion and curvature. ΔP2 has a steeper evolution for uniform velocity inlet scenario and the latter simulated point, namely $Q_t = Q_{t,max} = 1.1\ Q_p$, differs. In that operating condition, the inlet flow from the upper reservoir eases the flow momentum pointing to the turbine branch but only in the scenario with uniform velocity inlet. With a developed turbulent velocity profile coming from the pump, the mixing point with the discharge from the upper reservoir preserves higher pressure drops.

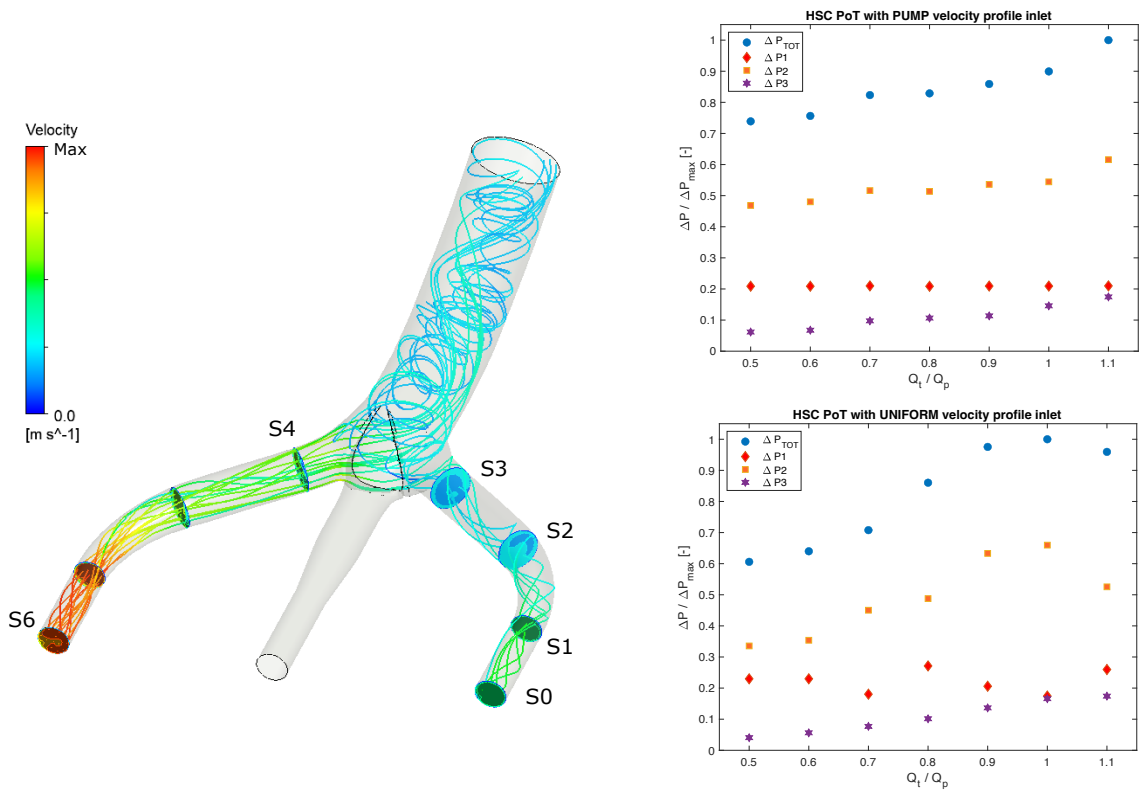

Fig. 6: CFD streamlines of case-1 at $Q_t = Q_{t,min} = 0.5\ Q_p$ and comparative plots of the relative pressure drops $\Delta P1$, $\Delta P2$, and $\Delta P3$.

### 5.1.2  HSC case-2

The solution of the case-2 with the imported pump velocity profile globally presents a lower pressure loss apart from the initial branches $\Delta P1$ due to the higher initial turbulence (Fig.7). The pressure losses $\Delta P3$ are depending on the turbine discharge and presents upward trend with the increase of the velocity in this branch. Fig. 8 shows the comparison of the surface streamlines and magnitude velocity for sections S3-S0 for the two different inlet velocity profiles. In the case with uniform velocity inlet, the flow structures in turbine branch are more complex by presenting two developed vortices entering the turbine branch, which justify the higher pressure losses for $\Delta P3$. $\Delta P2$ accounts for most of the total flow losses and it increases with the discharge of the turbine. However, with a uniformed velocity inlet, it records a favourable flow path with an incoming flow from the upper reservoir $Q = Q_{t,\ max}$. From a better insight of the flow field in the case-2, Fig.9 presents the surface streamlines at middle plane which illustrates that the sudden decrease of $\Delta P2$ from $Q_t = Q_p$ to $Q_t = 1.1 Q_p$ results from the abrupt change of the flow structure in the branch.

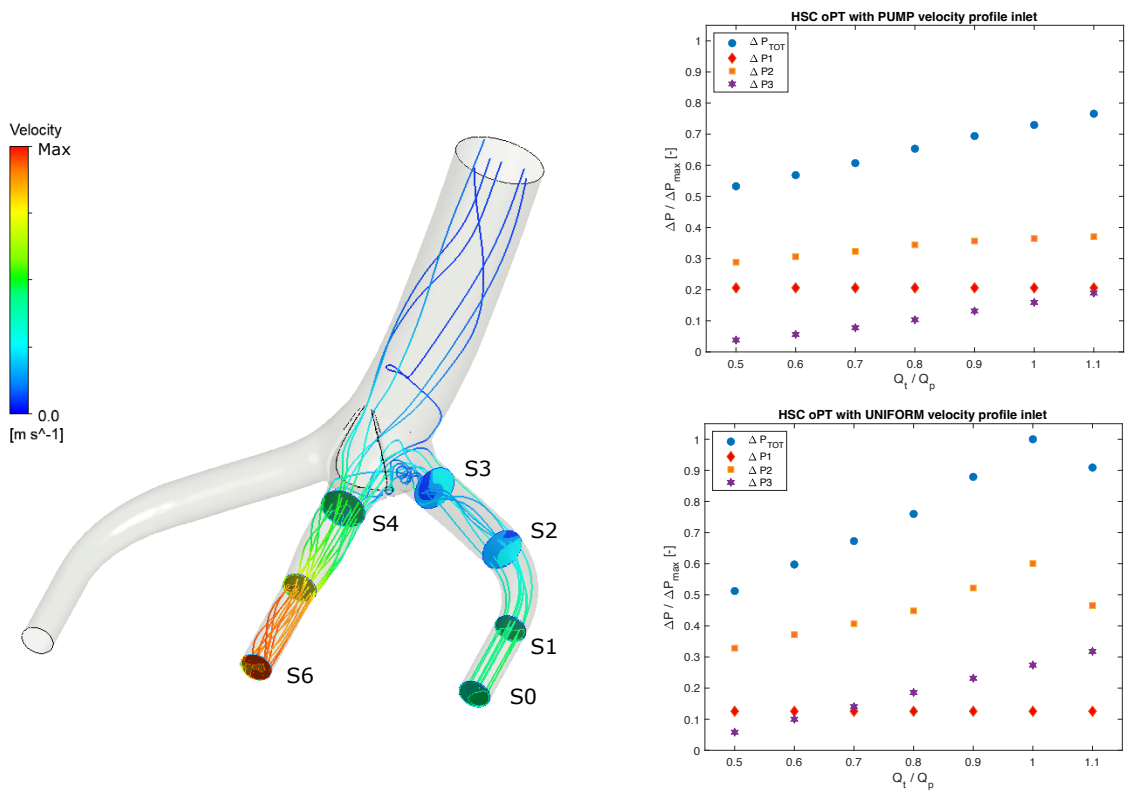

Fig. 7 CFD streamlines of case-2 at $Q_t = Q_{t,min} = 0.5\, Q_p$ and comparative plots of the relative pressure drops ΔP1, ΔP2, and ΔP3.

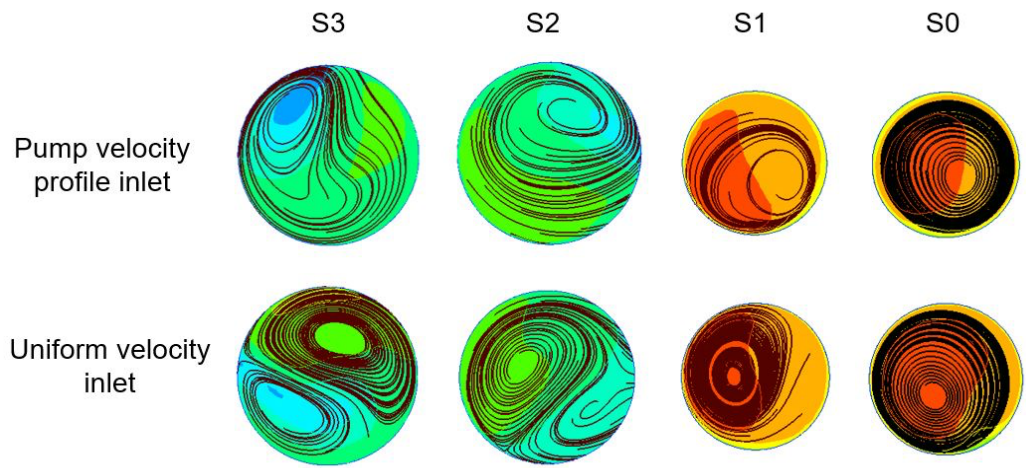

Fig.8 2D streamlines of case-2 at $Q_t = Q_p$ of sections S3 to S0

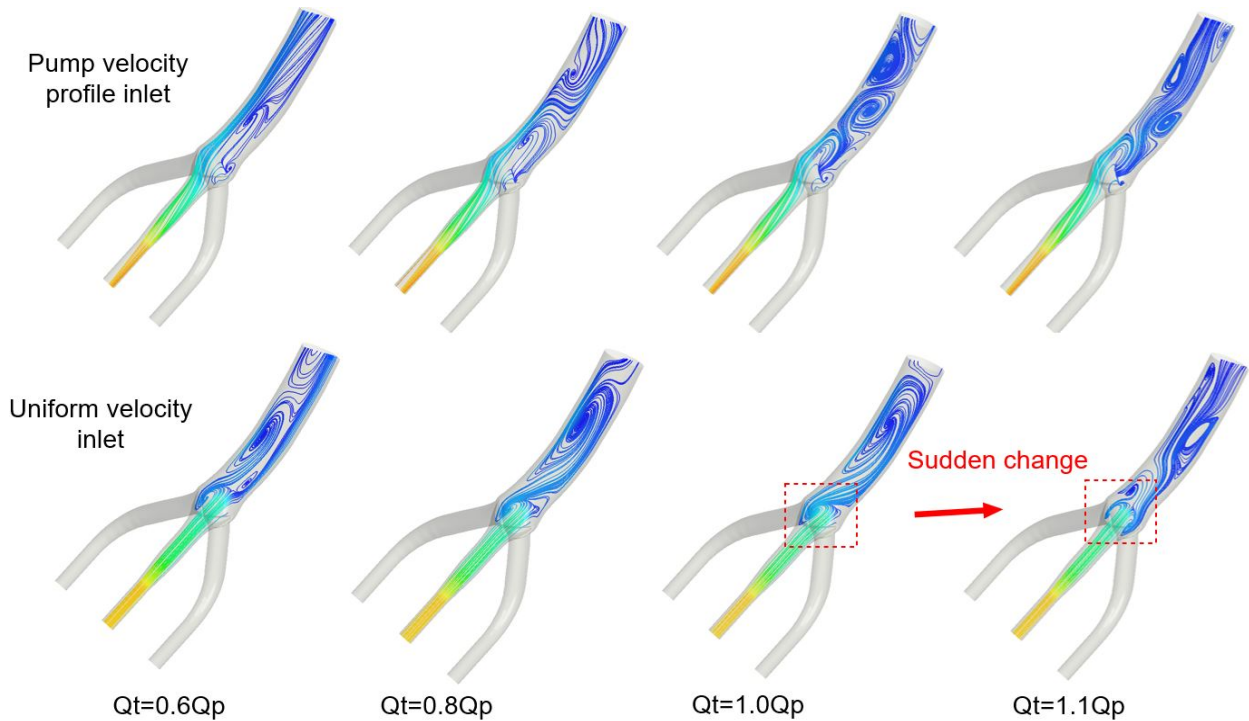

Fig.9 2D streamlines of case-2 at middle plane

## 6  Conclusions

This paper provides a methodology to numerically investigate the HSC feasibility in the look of assessing the off-design operating conditions and its potential. The study refers to an existing PHES plant equipped with three Francis pump-turbines that has not been designed for HSC application. The outcomes of this research propose a model expression for the power compensation of head loss due to the new operation based on 3D CFD numerical simulations to evaluate the flow condition in the PHES plant trifurcation. Under HSC operating conditions discussed in this paper, flow losses in penstock raises with turbine discharge rate and it reaches maximum when all pumping fluid feeds into turbine. Meanwhile, CFD results show losses in penstock are also affected by the level of inflow turbulence. The computation of the flow structure in the hydraulic system presents the correlation of total pressure losses and the flow patterns which indicates the loss mechanism of penstock under HSC conditions. Furthermore, the effect of the inlet velocity profile at the trifurcation results to be not negligible for complex mixing cross-sections, especially in studying different turbine discharge contribution at the trifurcation. From here, unsteady simulations will be needed to validate the results and to further investigate transient regimes that potentially could affect the system in HSC operations.

**Authors**

Dr. Alessandro MORABITO
EPFL, Lausanne - Technology Platform for Hydraulic Machines
Av. de Cour, 33/bis, 1007 Lausanne SWITZERLAND
E-mail Alessandro.morabito@epfl.ch

Chengshuo Wu
Zhejiang University, China
Zheda Road No.38, Hangzhou, China;
EPFL, Lausanne - Technology Platform for Hydraulic Machines
Av. de Cour, 33/bis, 1007 Lausanne SWITZERLAND
E-mail chengshuo.wu@epfl.ch

Stefano SIGALI
Enel Green Power Innovation
Via A. Pisano 120, 56122, Pisa, Italy
E-mail stefano.sigali@enel.com

Dr. Elena VAGNONI
EPFL, Lausanne
Technology Platform for Hydraulic Machines
Av. de Cour, 33/bis, 1007 Lausanne SWITZERLAND
E-mail Elena.Vagnoni@epfl.ch

**Alessandro Morabito** graduated in energy engineering in 2014 from Politecnico di Milano, Italy. He joined the Université Libre de Bruxelles, Belgium, for research activities in energy storage technologies and he received his PhD in Engineering Science and Technology (2017-2021). He is currently postdoc researcher at École Polytechnique Fédérale de Lausanne, Technology Platform for Hydraulic Machines for research in reaching flexibility capacity of pumped storage power plants.

**Chengshuo Wu** is a PhD researcher in fluid machinery of Zhejiang University and now she is participating in a one-year project in EPFL as a visiting PhD researcher. She focuses on the investigation of internal flow mechanism and transient characteristics of fluid machinery and has four-year experience in the high-efficiency and low-vibration pump design.

**Elena Vagnoni** graduated from Politecnico di Milano and Johns Hopkins University in 2013. She joined the EPFL Laboratory for Hydraulic Machines in 2014 to obtain her PhD in Mechanics, and she completed her doctoral studies in August 2018. Since 2020, she is scientist and lecturer at EPFL in the Technology Platform for Hydraulic Machines leading the research team and scientific projects. In particular, she performs and supervises the research activities related to the development of smart controls and methods for extending the hydropower plants flexibility.